\documentclass[amstex,aps,amsfonts,prsty]{article}

\usepackage{amssymb}
\usepackage{amsmath}
\usepackage{graphicx}
\usepackage{citesort}

\def \be{\begin{equation}}
\def \ee{\end{equation}}
\def \ba{\begin{array}{l}}
\def \Ba{\begin{array}{ll}}
\def \ea{\end{array}}
\def \bq{\begin{eqnarray}}
\def \eq{\end{eqnarray}}
\def \nn{\nonumber\\}
\def \lb{\label}
\def \ln{{\rm ln}}

\def \fr{\frac}

\def \b{\beta}
\def \d{\delta}
\def \D{\Delta}
\def \e{\epsilon}
\def \f{\phi}

\def \G{\Gamma}
\def \lm{\lambda}

\def \n{\nabla}
\def \p{\varphi}
\def \h{\chi}

\def \t{\tau}

\def \ol{\overline}

\def \[{\left[}
\def \]{\right]}
\def \({\left(}
\def \){\right)}

\def \R{R_{c}}
\def \T{T_{c}}

\def \I{\int d^{D}x}
\def \2{\frac{1}{2}}
\def \4{\frac{1}{4}}

\begin{document}

\begin{center}

{\Large \bf Off-perturbative states in disordered systems}

\vskip .2in

Vik.S.\ Dotsenko

\vskip .1in

LPTMC, Universit\'e Paris VI,  4 place Jussieu, 75252 Paris, France  

L.D.Landau Institute for Theoretical Physics, 
   117940 Moscow, Russia

\end{center}

\vskip .3in

\begin{abstract}
The systematic approach for the
off-perturbative calculations in disordered systems
is developed. The proposed scheme is applied for 
the random temperature and the random field ferromagnetic 
Ising models. It is shown that away from the critical point,
in the paramagnetic phase of the random temperature
model, and in the ferromagnetic phase of the random field
one, the free energy contains non-analytic
contributions which have the form of essential singularities.
It is demonstrated that these contributions 
appear due to localized in space instanton-like excitations.
\end{abstract}

\vskip .1in

{\bf Key words}:Quenched disorder, instantons, replicas, mean-field, 
non-analytic functions.

\vskip .3in

\section{Introduction}

In very simplified terms, studies of classical statistical 
systems 
involves two main domains: (1) investigation of the 
ground state, and (2) summation over fluctuations
around this ground state. Although formally
according to the definition of the partition function,
one has to perform summation over the whole 
configurational space of a system, in reality
it is never done. And it is not that we are doing 
something wrong. The point is that in most of the cases
only very limited part of the 
configurational space which is relevant for 
observable thermodynamics. Very often, it the 
question, what this "relevant part" is 
(which involves the choice of the so called  
"relevant variables"), which is the most difficult.
Studies of the systems containing quenched disorder,
in addition to the two items mentioned above,
involves the third one (although, technically,
very often it turns into the starting one),
which is the averaging of self-averaging quantities
over random parameters. Nevertheless, at a qualitative level,
the situation here remains the same:
only very limited part of the configurational space is
relevant for observable thermodynamics.

However, in some  statistical systems, 
besides the ground state 
another local minimum (or minima) could exists.
Let us consider extremely simplified
situation, schematically shown in Fig.1, 
when in addition to the
ground state the system has another local minimum
located in the configurational space
"far away" from the ground states, and separated from it 
by a big (compared to the temperature) energy barrier,
which, however, remains {\it finite} in the thermodynamic limit.
If we are dealing with the system which contains no
quenched disorder, then the thermodynamic contribution 
due to this another state with an exponential accuracy
will be simply of order of $\exp(-\b \D E)$ (provided 
$\D E \gg T$), where $\D E = E_{1}-E_{0}$. 
 The crucial point, however,
is that to get the above exponential contribution,
one has to {\it know} about existence of the other
local minimum, otherwise, its contribution would be just
missing. In other words, summing up the perturbation theory
around the ground state, and even
taking into account all non-linear terms of the
Hamiltonian, responsible for the existence of the other
local minimum, would not recover the contribution
$\exp(-\b \D E)$ of the other state, which is 
located "beyond barrier". It is these type of
contributions which are usually called "off-perturbative".

\begin{figure}[h]
\includegraphics[scale=0.30]{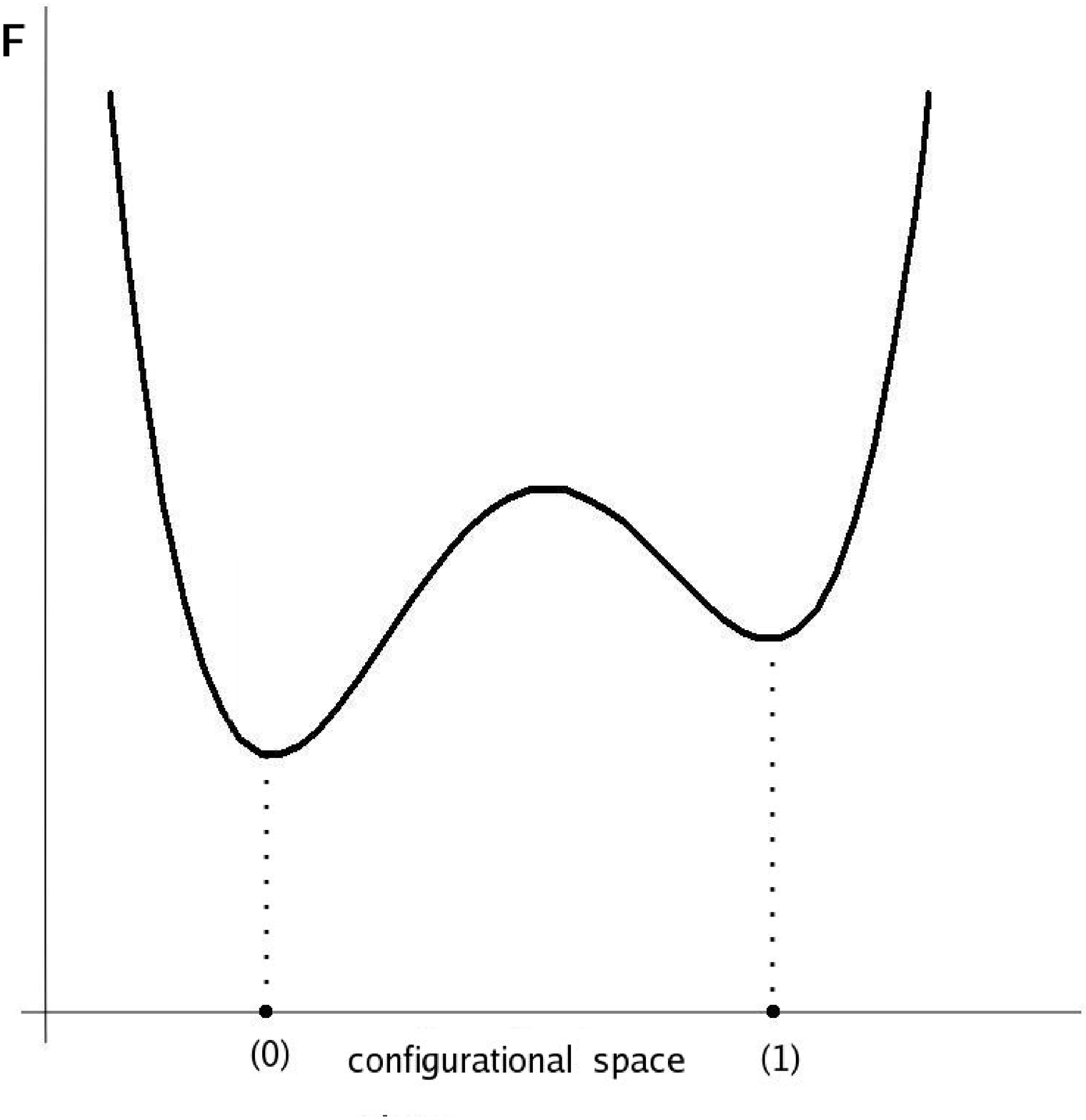}
\caption[]{Schematic structure of the free energy
landscape with two well separated  
thermodynamically relevant valleys}
\end{figure}

In the studies of the effects produced by the 
quenched disorder, conditionally, one can distinguish 
two main domains of research: strongly disordered
systems, like spin-glasses (where the disorder
in the dominant factor), and the systems containing 
some kind of weak disorder which is supposed not
to destroy the ground states properties of the 
corresponding pure system. 
Traditionally, magnetic statistical systems
containing weak disorder, such as
random bond ferromagnetic Ising models,
are studied focusing mainly on modifications
introduced into their
critical properties at the phase transition 
point\cite{harris,RG-disorder}.
In fact, as was pointed out by Griffith
many years ago\cite{grif}, 
modification of the critical behavior,
is not the only qualitative physical 
phenomenon which can be produced here.

Let us come back to the example shown in Fig.1.
The presence of weak quenched disorder here, 
provided it does not ruin the global structure 
of the phase space, would just 
require supplementary averaging of
the above exponential
factor, $\exp(-\b \D E)$ (since both
the energy of the ground state $E_{0}$ and the
energy of the excited state $E_{1}$ are now the 
functions of the disorder parameters), 
but qualitatively, it would not modify
the situation too much. 
Completely different and new physical phenomena
comes into play when the structure of
the phase space similar to that shown in Fig.1 
{\it is created} by the presence of randomness.
In other words, this is the situation when
weak quenched disorder, although it 
does not modify the ground state of the system,
creates something completely new
(absent in the corresponding pure system),
namely, the local minima states, somewhere
at the periphery of the phase space,
"far away" from the ground state of the system.

According to the original observation by
Griffiths\cite{grif} and later 
studies\cite{grif-stat,zeros-cardy}, the presence of
such off-perturbative states in the disordered
ferromagnetic Ising model makes its free energy
to be {\it non-analytic} function of the 
external magnetic field $h$ in a whole temperature
interval above the ferromagnetic phase transition
point. Moreover, at least in some cases,
this non-analyticity has the form of
essential singularity in the limit $h\to 0$,
\cite{grif-stat,grif-dos}.
It has to be stressed that all such contributions 
are just missing in the
traditional (perturbative) RG treatment of the
problem\cite{RG-disorder}.
In more spectacular way the off-perturbative effects 
manifest themselves in the dynamical properties,
producing the slowering down
of the relaxation processes\cite{grif-dynam},
as well as in the quantum systems 
(see e.g. \cite{grif-quant} and references therein).

Although at a qualitative level the origin of 
the off-perturbative contributions
is more or less clear, their technical
implementation, namely, the derivation
of e.g. the non-analytic part of the free energy,
turns out to be rather tricky problem. 
Usually, analytic calculations in disordered systems
are performed in terms of the replica method.
Many years ago Parisi has suggested\cite{grif-parisi}, 
that the presence of additional
local minima configurations in weakly disordered systems
is related, in the replica approach,
to the existence of localized in space
and breaking replica symmetry instanton-like excitations
(translation invariance and replica symmetry is
recovered by taking into account all possible excitations
of this kind). Later on there were several attempts
of concrete realization of this idea for
the random temperature\cite{zeros-cardy} and
the random field\cite{RF-grif-dos1,RF-grif-dos2} Ising models
where it has been demonstrated that the 
corresponding saddle-point equations may indeed have
instanton-like solutions. Next step has been done
when the systematic method of summation over 
all such type of solutions, breaking symmetry
in the replica {\it vector} order parameter, has been 
developed\cite{vrsb,dos-book}. In terms of this method 
the explicit form of the off-perturbative contributions  
in the random temperature Ising model 
has been derived\cite{grif-inst}. 

In the present paper (following the recent study
the original Griffith problem\cite{grif-dos}) 
the systematic approach for the
off-perturbative calculations and its relation
with the method of the vector replica 
symmetry breaking\cite{vrsb,dos-book} is formulated
(section II). Then, proposed scheme is applied for 
the random temperature (section III) and the random field 
(section IV) ferromagnetic Ising models.
It is shown that in both systems at temperatures
away from the critical point, namely,
in the paramagnetic phase of the random temperature
model, and in the ferromagnetic phase of the random field
one, the free energy contains non-analytic
contributions which (as the functions of the 
disorder parameters) have the form of essential singularities.
It is demonstrated that these contributions 
appear due to localized in space instanton-like
configurations. Physical discussion of the obtained results
is given in Section V.

\section{General scheme of calculations}

Let us consider a general (continuous) 
$D$-dimensional random system 
described by a Hamiltonian $H\[\f({\bf x}); \xi({\bf x})\]$, 
where $\f({\bf x})$ is a field which defines
the microscopic state of the system, and 
$\xi({\bf x})$ are quenched random parameters.
Let us suppose that in addition to the
ground state, this system has another thermodynamically 
relevant (Griffith) region of the 
configurational space located "far away"
from the ground state and separated from it 
by a finite barrier of the free energy (see Fig.1).
In other words, it is  {\it supposed} that the partition 
function (of a given sample) can be represented in
the form of two separate contributions:
\be
\lb{st2.1}
Z  = \int {\cal D}\f({\bf x}) \mbox{\large $e$}^{-\b H} \; = \; 
  \mbox{\large $e$}^{-\b F_{0}} + \mbox{\large $e$}^{-\b F_{1}} 
 \; \equiv \;Z_{0} \; + \; Z_{1}
\ee
where $F_{0}$ is the contribution coming 
from the vicinity of the ground state,
and $F_{1}$  is the contribution of the 
Griffiths region. Then, for the averaged
over disorder total free energy we find:
\be
\lb{st2.2}
{\cal F} = -\fr{1}{\b} \ol{\ln Z} 
= \ol{F_{0}} - \fr{1}{\b} 
\ol{\ln\[1 + Z_{1} Z_{0}^{-1}\]}
\ee
The second term in the above equation, 
which will be denoted by $F_{G}$, 
can be represented in the form of the series:
\be
\lb{st2.3}
F_{G} = - \fr{1}{\b} \sum_{m=1}^{\infty}
\fr{(-1)^{m-1}}{m} \ol{Z_{1}^{m} Z_{0}^{-m}}
= - \fr{1}{\b} \lim_{n\to 0} \sum_{m=1}^{\infty}
\fr{(-1)^{m-1}}{m} Z_{n}(m)
\ee	      
where
\be
\lb{st2.4}
Z_{n}(m) = \prod_{b=1}^{m} \int {\cal D}\f^{(1)}_{b}
         \prod_{c=1}^{n-m} \int {\cal D}\f^{(0)}_{c} \; 
\mbox{\Large $e$}^{ -\b H_{n}\[\f^{(1)}_{1},...,\f^{(1)}_{m}, 
           \f^{(0)}_{1},...,\f^{(0)}_{n-m}\]}    
\ee
is the replica partition function
($H_{n}\[{\boldsymbol\phi}\]$ is the corresponding
replica Hamiltonian), 
in which the replica symmetry in the $n$-component 
vector field $\f_{a}$ ($a=1,...,n$) is assumed to be 
broken. Namely, it is supposed that the saddle-point
equations
\be
\lb{st2.5}
\fr{\d H_{n}\[{\boldsymbol\phi}\]}{\d \f_{a}({\bf x})} \; = \; 0
\;, \; \; \; \; (a = 1, ..., n)
\ee
have non-trivial solutions with the RSB structure
\be
\lb{st2.6}
\f_{a}^{*}({\bf x}) = \left\{ \begin{array}{ll}
                 \f_{1}({\bf x})   & \mbox{for $a = 1, ..., m$}
		 \\
		 \\
                 \f_{0}({\bf x})   & \mbox{for $a = m+1, ..., n$}
                            \end{array}
                            \right.
\ee
with $\f_{1}({\bf x}) \not= \f_{0}({\bf x})$, so that the integration
in the above partition function, eq.(\ref{st2.4}), goes over
fluctuations in the vicinity of these components:
\bq
\lb{st2.7}
\f^{(1)}_{b}({\bf x}) &=& \f_{1}({\bf x}) + \p_{b}({\bf x}) ,  \; \; \; \; 
          (b =  1, ..., m)
\nn
\f^{(0)}_{c}({\bf x}) &=& \f_{0}({\bf x}) + \h_{c}({\bf x}) , \; \; \; \; 
          (c =  1, ..., n-m)
\eq

It should be stressed that to be thermodynamically relevant,
the RSB saddle-point solutions, eq.(\ref{st2.6}), 
should satisfy the following tree crucial conditions:

 (1) the solutions should be {\it localized} in space, 
so that they are characterized by  {\it finite}
space sizes $R(m)$; in this case the partition 
function, eq.(\ref{st2.4}), will be proportional to
the entropy factor $V/R^{D}(m)$ 
(where $V$ is the volume of the system), and
the corresponding free energy contribution 
$F_{G}$, eq.(\ref{st2.3}), 
will be extensive quantity;
 
 (2) they should have {\it finite} energies 
$E(m) = H_{n}\[{\boldsymbol\f}^{*}\]$;

 (3) the corresponding Hessian matrix of these solutions 
should have all eigenvalues positive. 

Thus, in the systematic calculations one should
find all saddle-point RSB solutions $\f_{a}^{*}({\bf x})$,
eq.(\ref{st2.6}), (satisfying the above three requirements), 
after that one has to compute
their energies $E(m)$ (for $n\to 0$), 
next one has to
integrate over the fluctuations in the vicinity of 
these solutions, and finally one has to 
sum up the series

\be
\lb{st2.8}
F_{G} = -\fr{V}{\b} \sum_{m=1}^{\infty} \fr{(-1)^{m-1}}{m}  
   R^{-D}(m) \; \(\b \det \hat T\)^{-1/2}_{n=0} \; 
\mbox{\Large $e$}^{-\b E(m)}
\ee
where $\hat T$ is the ($n\times n$) matrix
\be
\lb{st2.9}
T_{aa'} \; = \; 
\fr{\d^{2} H\[{\boldsymbol\f}\]}{\d\f_{a} \d\f_{a'}}\Bigl|_{
{\boldsymbol\f}={\boldsymbol\f}^{*}}
\ee
Note that in the present approach 
the procedure of analytic continuation  
$n \to 0$ is quite similar to that in the usual 
replica theory \cite{MPV-book}: whenever the parameter
$n$ becomes an algebraic factor (and not the summation
parameter, or the matrix size, etc.), it can safely
be set to zero right away.

\vspace{5mm}

The above scheme of calculations can be easily 
generalized for an arbitrary number of the
Griffiths regions.
For example, let us
consider the situation which is qualitatively
represented in Fig.2, when in addition to the 
ground state, the system has {\it two} thermodynamically
relevant Griffiths states. 
In this case instead of eq.(\ref{st2.1}) we will have
\begin{center}
\begin{figure}[h]
\includegraphics[scale=0.30]{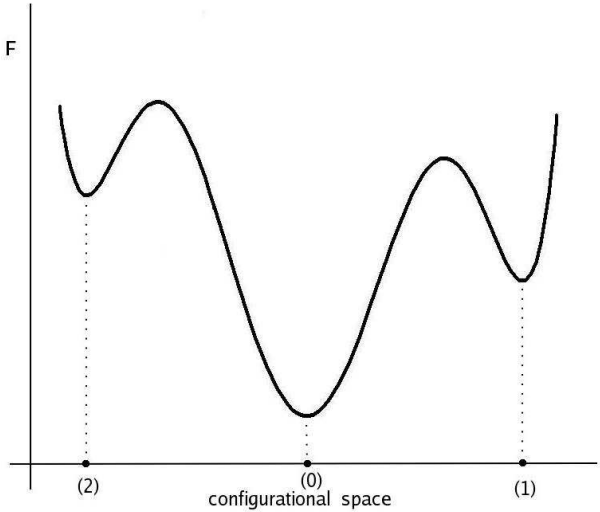}
\caption[]{Schematic structure of the free energy
landscape with three well separated  
thermodynamically relevant valleys}
  \label{fig2}
\end{figure}
\end{center}

\be
\lb{st2.10}
Z  = \int {\cal D}\f(x) \mbox{\large $e$}^{-\b H} \; = \; 
  \mbox{\large $e$}^{-\b F_{0}} + \mbox{\large $e$}^{-\b F_{1}} 
  + \mbox{\large $e$}^{-\b F_{2}}
  \; \equiv \; Z_{0} \; + \; Z_{1} \; + \; Z_{2}
\ee
and correspondingly, instead of eq.(\ref{st2.3})
we find

\bq
\lb{st2.11}
F_{G} &=& -\fr{1}{\b} 
\ol{\ln\[1 + Z_{1} Z_{0}^{-1} + Z_{2} Z_{0}^{-1}\]} \; = \;
- \; \fr{1}{\b} \sum_{m=1}^{\infty}
\fr{(-1)^{m-1}}{m} 
\sum_{k=0}^{m} \fr{m!}{k! (m-k)!} \; 
\ol{\(Z_{1}^{k} Z_{2}^{m-k}Z_{0}^{-m}\)}
\nn
\nn
&=& -\fr{1}{\b} \lim_{n\to 0} \sum_{m=1}^{\infty}
\fr{(-1)^{m-1}}{m} 
\sum_{k=0}^{m} \fr{m!}{k! (m-k)!}  \; 
Z_{n}(k,m)
\eq	      
Here, in the replica partition function

\be
\lb{st2.12}
Z_{n}(k,m) = \prod_{b=1}^{k} \int {\cal D}\f^{(1)}_{b}
            \prod_{c=1}^{m-k} \int {\cal D}\f^{(2)}_{c}
            \prod_{d=1}^{n-m} \int {\cal D}\f^{(0)} \;    
	    \mbox{\Large $e$}^{-\b H_{n}
	    \[{\boldsymbol\f}^{(1)},
              {\boldsymbol\f}^{(2)},
	      {\boldsymbol\f}^{(0)}\]}
\ee
the integration is supposed to be performed  
in the vicinity of the saddle-point replica vector
\be
\lb{st2.13}
\f_{a}^{*}({\bf x}) = \left\{ \begin{array}{ll}
                 \f_{1}({\bf x}),  & \mbox{for $a = 1, ..., k$}
		 \\
		 \\
                 \f_{2}({\bf x}),  & \mbox{for $a = k+1, ..., m$}
		 \\
		 \\
                 \f_{0}({\bf x}),  & \mbox{for $a = m+1, ..., n$}
                            \end{array}
                            \right.
\ee
(where $\f_{1}({\bf x}) \not= \f_{2}({\bf x}) \not= \f_{0}({\bf x})$)
which is the solution of the saddle-point 
equations (\ref{st2.5}).
Finally, for the Griffiths free energy contribution, 
instead of eq.(\ref{st2.8}) one obtain

\be
\lb{st2.14}
F_{G} = -V \sum_{m=1}^{\infty} \fr{(-1)^{m-1}}{\b m}   
\sum_{k=0}^{m} \fr{m!}{k! (m-k)!}  \; R^{-D} \(\b \det\hat T\)^{-1/2}_{n=0}
\mbox{\Large $e$}^{-\b E(k,m)}
\ee
where  
$E(k,m)= H_{n\to 0}\[{\boldsymbol\f}^{*}\]$ 
is the energy of a given solution,
eq.(\ref{st2.13}), and $\hat T$ is the Hessian 
matrix, eq.(\ref{st2.9}).

\vspace{5mm}

It is interesting to note that one can arrive 
to the same representations for the 
off-perturbative free energy contributions,
eq.(\ref{st2.8}) or eq.(\ref{st2.14}), following
the so called vector replica
symmetry breaking scheme\cite{vrsb,dos-book}.
The  starting point here is 
the  standard replica definition for the averaged over 
disorder free energy,
\be
\lb{st2.15}
{\cal F} = -\fr{1}{\b} \lim_{n\to 0} \fr{\ol{Z^{n}} - 1}{n}
\ee
where the replica partition function 
$\ol{Z^{n}}\equiv Z_{n}$ is formally
defined by the integration over all 
configurational space:
\be
\lb{st2.16}
Z_{n} \; = \; \prod_{a=1}^{n} \int {\cal D}\f_{a} \; 
\mbox{\Large $e$}^{ -\b H_{n}\[\f_{1},...,\f_{n}\]}    
\ee
Now, let us suppose that in addition to the usual 
replica symmetric (RS) ground state configuration,
the saddle-point equations (\ref{st2.5}) have
another types of solutions, which are {\it well separated} 
in the configurational space from the RS state.
In this case (again, denoting their contributions 
by the label "G") the replica partition function,
eq.(\ref{st2.16}), can be decomposed into two parts:
\be
\lb{st2.17}
Z_{n} \;  = \; Z_{RS} + Z_{G}
\ee
Here $Z_{RS}$ contains all "routine" perturbative contributions 
in the vicinity of the ground state, and, as usual, 
(in the limit $n\to 0$) this partition function can be
represented in the form:
\be
\lb{st2.18}
Z_{RS} \; = \; \mbox{\Large $e$}^{-\b n F_{0}}
\ee
Thus, according to eq.(\ref{st2.15})
for the total free energy we get:
\be
\lb{st2.19}
{\cal F} \; = \; F_{0} + F_{G}
\ee
where 

\be
\lb{st2.20}
F_{G} = - \lim_{n\to 0} \fr{1}{\b n} Z_{G}
\ee
contains all {\it non-replica-symmetric} contributions (if any).
As an example, let us suppose that the saddle-point 
eqs.(\ref{st2.5}) have non-trivial solutions with
the three groups structure like in eq.(\ref{st2.13}).
Moreover, let us suppose that these solutions possess 
three crucial properties: 
(1) they are localized in space and characterized by
finite spatial sizes $R_{n}(m,k)$; 
(2) they have finite energies $E_{n}(k,m)$; 
and (3) their Hessian matrices $T_{ab}$, eq.(\ref{st2.9}), 
have all the eigenvalues positive. 
Than taking into account all possible permutations of the
tree replica vector components the above free energy 
$F_{G}$ can be represented in the form
\be 
\lb{st2.21}
F_{G} = - \lim_{n\to 0} \fr{1}{\b n} 
       \sum_{m=1}^{n} \fr{n!}{m! (n-m)!} 
       \sum_{k=0}^{m} \fr{m!}{k! (m-k)!} 
       \fr{V}{R_{n}^{D}(k,m)} \(\b\det\hat T\)^{-1/2} 
       \mbox{\Large $e$}^{-\b E_{n}(k,m)}
\ee
To perform the analytic continuation $n \to 0$ in the above expression
the parameter $n$ must enter as an algebraic factor,
and not as the parameter of summation. This can be 
achieved if we represent the above
series in the following way:
\be
\lb{st2.22}
F_{G} = -\lim_{n\to 0}\fr{1}{\b n}
       \sum_{m=1}^{\infty} \fr{\G(n+1)}{\G(m+1)\G(n-m +1)} 
       \sum_{k=0}^{m} \fr{m!}{k! (m-k)!} 
       \fr{V}{R_{n}^{D}(k,m)} \(\b\det\hat T\)^{-1/2} 
       \mbox{\Large $e$}^{-\b E_{n}(k,m)} 
\ee
Here the summation over $m$ is extended beyond 
$m=n$ limit since the
gamma function $\G(z)$is equal to infinity both at $z=0$
and at all negative integers.
Now using the relation:
\be
\lb{st2.23}
\G(-z) = -\fr{\pi}{z\G(z)\sin(\pi z)}
\ee 
and referring to "good" analytical properties 
of the Gamma functions,
we can perform the analytic continuation $n \to 0$:
\bq
\lb{st2.24}
\fr{\G(n+1)}{\G(m+1)\G(n-m+1)} &=&
\fr{\G(n+1)}{\G(m+1)\G[-(m-1-n)]}
\nn
\nn
&=& - 
\fr{\G(n+1)(m-1-n)\G(m-1-n)\sin[\pi(m-1)-\pi n]}{\pi \G(m+1)}
\nn
\nn
&=&
\fr{\sin(\pi n)}{\pi} \; (-1)^{m-1} \;
\fr{\G(n+1)\G(m-n)}{\G(m+1)}
\Bigl|_{n\to 0} 
\nn
\nn
&\simeq&
n \; \fr{(-1)^{m-1}}{m}
\eq
Substituting this into eq.(\ref{st2.22}) we obtain eq.(\ref{st2.14})
(where $R(k,m) \equiv R_{n=0}(k,m)$ and 
$E(k,m) \equiv E_{n=0}(k,m)$)

Now, in the next two sections I am going to demonstrate
how the above general scheme works in the concrete
cases of the random temperature and the random field
ferromagnetic Ising models.

\section{Random temperature Ising model}

Let us consider weakly disordered $D$-dimensional
Ising model described by the continuous
Ginzburg-Landau Hamiltonian:
\be
\lb{st3.1}
H = \I \Biggl[ \2 \(\n\f({\bf x})\)^{2}  
  + \2 \(\t - \d\t({\bf x})\) \f^{2}({\bf x})   
  + \4 g \f^{4}({\bf x}) \Biggr] 
\ee  
The disorder is modeled here by a
random function $\t({\bf x})$ which is described by the 
Gaussian distribution,
\be
\lb{st3.2}
P[\d\t] = p_{0} \exp \Biggl( -\frac{1}{4u}\I (\d\t({\bf x}))^{2} \Biggr) \ ,
\ee
where $u$ is the small parameter which describes 
the strength of the disorder, and $p_{0}$ is the 
normalization constant. We are going to consider 
this system in the paramagnetic phase away from
the critical point, so that the reduced temperature 
parameter $\t$ will be taken to be positive 
and not too small (it will be demonstrated below that
in dimensions $D < 4$ the limitation on the value of $\t$ 
is given by the usual Ginzburg-Landau condition,
$\t \gg g^ {2/(4-D)}$).

Weakly disordered systems described in terms
of the continuous Ginzburg-Landau Hamiltonian have been usually 
studied in the framework of the renormalization-group (RG),
(perturbative)  treatment. In this approach one 
is able to perform the systematic integration  
over all fluctuations at the background of the 
homogeneous state
up to the scales of the correlation length $R_{c}(\t)$.
In some cases, this makes possible to derive the leading 
singularities of the thermodynamical functions in
the critical point, at $\t\to 0$, where $\R(\t)$
diverges\cite{RG-disorder}. 

However, considering paramagnetic phase of this system,
intuitively, it is clear that the  contributions of 
non-homogeneous local minima configurations 
at scales bigger that the correlation
length, which exist due to rare localized in space
"ferromagnetic islands" with negative effective value of the 
local temperature parameter $(\t - \d\t)$ are missing in the
traditional RG treatment of the problem. To what extend 
these states are relevant for the critical behavior
(at $\t\to 0$) is still unclear\cite{critical-rsb}.
In the present study , however, 
I am going to address much more simple question:
what is the explicit form of the free energy contributions
due to such off-perturbative states away from $\T$,
(at $\t \gg \t_{g}$).

In fact, at purely heuristic level, it is not so difficult
to estimate form of these contributions. Let us consider
the spatial island of the linear size  $L$ characterized by
the typical value of the "local temperature" 
$(\t - \d \t) = - \xi < 0$
Its probability is exponentially small,
\be
\lb{stA}
{\cal P}\[L, \xi\] \; \sim \; 
\exp \Biggl( -\frac{(\t + \xi)^{2}}{4u} \; L^{D} \Biggr) \ ,
\ee
and therefore such islands are well separated from each
other and can be considered as non-interacting.

It has to be noted that the island with small
(negative) value of the  local temperature parameter
$-\xi$ can be
characterized by the mean-field
"up" and "down" states only if its size is
bigger than its local correlation length
$\R(\xi) \sim \xi^{-1/2}$. Thus, the
contribution to the free energy coming from the local 
ferromagnetic states of such islands with the exponential
accuracy can be estimated by their
probability:

\be
\lb{stB}
F_{G} \; \sim \; \int_{0}^{\infty} d\xi 
         \int_{\R(\xi)}^{\infty} dL
         \exp\[ -\frac{1}{4u} (\t + \xi)^{2} \; L^{D} \] \;
\sim \; \int_{0}^{\infty} d\xi 
     \exp\[ -(const)\frac{(\t + \xi)^{2}}{u} \xi^{-D/2} \]
\ee
Here in the integration over $\xi$ the leading contribution
comes from the vicinity of the saddle-point value
 \be
\lb{stC}
\xi_{*} \; = \; \fr{D}{4 - D} \; \t
\ee
(which is positive in dimensions $D<4$, 
and $\xi_{*} \gg \t_{g}$ provided $\t \gg \t_{g}$).
In this way we obtain
the following estimate for the off-perturbative
contributions coming from rare locally
ferromagnetic islands:
\be
\lb{stD}
F_{G}  \; \sim \; \exp\[ -(const)\frac{\t^{(4-D)/2}}{u} \]
\ee

Now let us consider how this result 
(including the value of the 
$(const)$ factor) can be derived analytically 
in terms of the systematic approach developed
in the previous section.
This derivation has been already
reported elsewhere\cite{grif-dos,grif-inst}.
Here I am going to give some more details about 
the corresponding replica instanton solutions, but
in general this section can be considered just 
as a "warming up" exercise before passing to 
more difficult calculations for the random field
model considered in the next section.

Performing the standard Gaussian integration
over random parameters $\d\t({\bf x})$, for the replica partition
function one gets

\be
\lb{st3.3}
Z_{n} = \prod_{a=1}^{n} \int {\cal D}\f_{a}(x) \;    
	    \mbox{\Large $e$}^{-\b H_{n}
	    \[{\boldsymbol\f}\]}
\ee
where 
\be
\lb{st3.4}
H_{n}\[{\boldsymbol\f}\] \; = \;
 \I \Biggl[ \2 \sum_{a=1}^{n}\(\n\f_{a}\)^{2} 
      + \2 \t \sum_{a=1}^{n} \f_{a}^{2} 
      + \4 g \sum_{a=1}^{n} \f_{a}^{4} 
      - \4 u \sum_{a,b=1}^{n}  \f_{a}^{2} \f_{b}^{2}
   \Biggr]
\ee
is the corresponding replica Hamiltonian.
The saddle-point configurations of the fields
$\f_{a}({\bf x})$ are defined by the equations
\be
\lb{st3.5}
-\D\f_{a}({\bf x}) + \t \f_{a}({\bf x}) + g \f_{a}^{3}({\bf x}) 
-u \f_{a}({\bf x}) \(\sum_{b=1}^{n} \f_{b}^{2}({\bf x})\) = 0
\ee  
Below we are going to demonstrate that besides the 
trivial solution $\f_{a}({\bf x}) = 0$ these equations
have non-trivial localized in space
instanton-like solutions with the RSB 
(two groups) structure:

\be
\lb{st3.6}
\f_{a}^{*}({\bf x}) = \left\{ \begin{array}{ll}
                 \f_{1}({\bf x})
		  & \mbox{for $a = 1, ..., m$}
		 \\
		 \\
                 0  & \mbox{for $a = m+1, ..., n$}
                            \end{array}
                            \right.
\ee
Substituting this anzats into the saddle-point eqs(\ref{st3.5})
and into the Hamiltonian, eq.(\ref{st3.4}), we find that
(in the limit $n\to 0$) the instanton configuration 
$\f_{1}(x)$ is defined by the equation
\be
\lb{st3.7}
-\D\f_{1}({\bf x}) + \t \f_{1}({\bf x}) - \lm(m) \f_{1}({\bf x})^{3} = 0
\ee
which is controlled by the parameter
\be
\lb{st3.8}
\lm(m) = u m  - g
\ee
and the energy of this configuration is
\be
\lb{st3.9}
E(m) = m \I \Biggl[\2 (\n\f_{1})^{2}  +
	       \2 \t \f_{1}^{2} - \4 \lm(m) \f_{1}^{4} \Biggr] 
\ee
In what follows the parameter $\lm(m)$ will be assumed 
to be {\it positive}. In other words, the solution, which 
we are going to derived below, exists only for $m$ such that 
$m \; > \; [g/u]$ (where $[...]$ denotes the integer part).
It has to be noted that one should not be confused 
by the "wrong" sign of the coupling
$\f^{4}$ term in the above equations.
In fact, it can be shown that the integration over the 
replica fluctuations around
considered solution in the limit $n\to 0$ yields the Hessian 
matrix which has all the eigenvalues positive
(this is the usual situation for the replica theory, where
 the minima of the physical quantities in the limit $n\to 0$
turns into maxima of the corresponding replica 
quantities\cite{MPV-book,dos-book}).

Rescaling the fields,
\be
\lb{st3.10}
\f_{1}({\bf x}) = \sqrt{\fr{\t}{\lm(m)}} \psi({\bf x} \sqrt{\t})
\ee
and introducing ${\bf z} \equiv {\bf x} \sqrt{\t}$, 
instead of eq.(\ref{st3.7}) one get the differential 
equation which contains no parameters:
\be
\lb{st3.11}
-\D \psi({\bf z}) + \psi({\bf z}) - \psi^{3}({\bf z}) = 0
\ee
Correspondingly, for the energy of this configuration, 
eq.(\ref{st3.9}), one obtains:
\be
\lb{st3.12}
E(m) = \fr{m}{um - g} \t^{(4-D)/2} E_{0}(D)
\ee
where the quantity $E_{0}(D)$ depends only on the dimensionality
of the system:
\be
\lb{st3.13}
E_{0}(D) = \int d^{D}z \[ \2 (\n\psi({\bf z}))^{2} + 
        \2 \psi^{2}({\bf z}) - \4 \psi^{4}({\bf z}) \]
\ee
It can be shown (see e.g. \cite{inst})
that in dimensions $D < 4$ 
eq.(\ref{st3.11}) has the smooth (with $\psi'(0) =0$)
spherically symmetric instanton-like solution $\psi(r)$ 
(where $r = |{\bf z}|$) such that:

\bq
\lb{st3.14}
\psi(r \leq 1) &\sim& \psi(r=0) \equiv \psi_{0} \sim 1, 
\nn
\nn
\psi(r \gg  1) &\sim& \mbox{\Large $e$}^{-r} \to 0. 
\eq
The energy $E_{0}(D)$ 
of this  solution is a finite and {\it positive} number.
In particular, in dimensions
$D = 3$, $\psi_{0} \simeq 4.34$ and $E_{0} \simeq 18.9$
(see Fig.3). 
\begin{center}
\begin{figure}[h]
\includegraphics[scale=0.40]{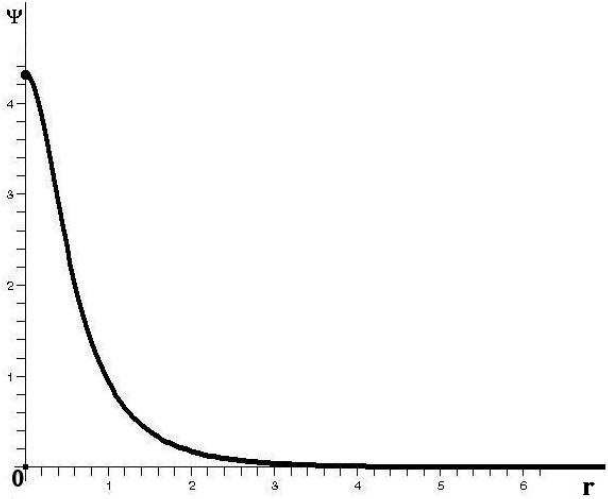}
\caption[]{The instanton solution $\psi(r)$ of eq.(\ref{st3.11})
in dimensions $D=3$}
  \label{fig3}
\end{figure}
\end{center}
As the dimension parameter $D$ approaches
the upper critical dimensionality $D_{c} = 4$,
from below the value of the field at the origin 
$\psi_{0}(D)$ 
tend to infinity (see Fig.4), while the energies 
of the corresponding
instanton configurations $E_{0}(D)$ 
approach the finite universal value 
$E_{0}(D\to 4) = E_{*} \simeq 26.3$. 
Above dimensions $D=4$ eq.(\ref{st3.11}) has 
no smooth instanton-like solutions.
In other words, described in terms 
of the dimensions 
parameter $D$, when passing the critical value
$D_{c}=4$ from below,  the instanton solution 
disappears in the discontinuous way.

\begin{center}
\begin{figure}[h]
\includegraphics[scale=0.40]{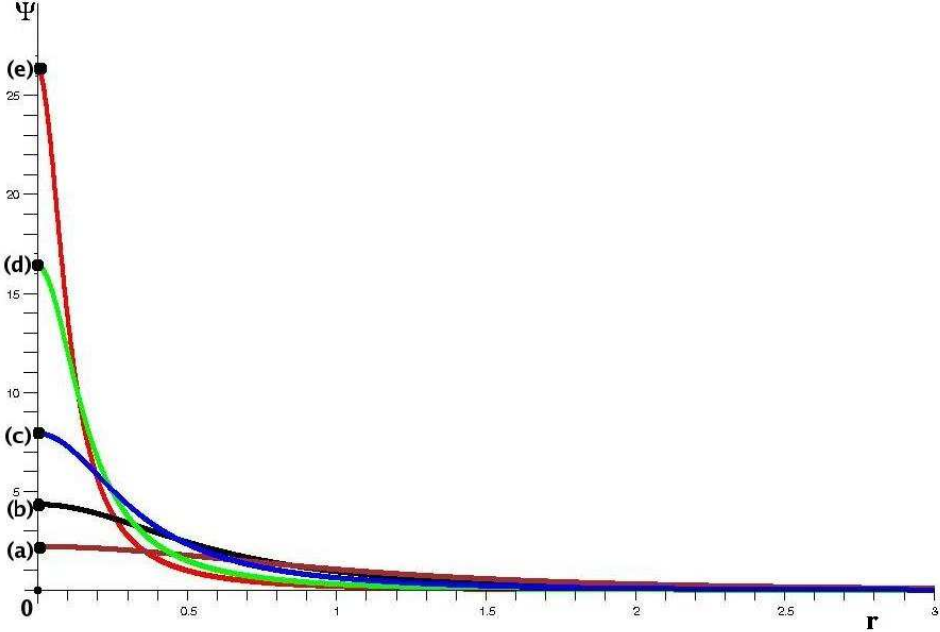}
\caption[]{The instanton solutions $\psi(r)$ of eq.(\ref{st3.11})
in various dimensions: 
(a) $D=2$,   $\; \psi_{0}\simeq 2.21$;
(b) $D=3$,   $\; \psi_{0}\simeq 4.34$;
(c) $D=3.5$, $\; \psi_{0}\simeq 7.92$;
(d) $D=3.8$, $\; \psi_{0}\simeq 16.4$;
(e) $D=3.9$, $\; \psi_{0}\simeq 26.4$;}
  \label{fig4}
\end{figure}
\end{center}

Note that according to the rescaling, eq.(\ref{st3.10}), 
the size of these instanton solutions 
in terms of the original fields $\f_{1}({\bf x})$
is $R_{c}(\t) = \t^{-1/2}$ (which is the usual correlation 
length of the Ginsburg-Landau theory) 
and it does not depends on $m$. 
Note also that due to obvious symmetry property
$\f_{a} \to -\f_{a}$ 
of the original saddle-point eqs.(\ref{st3.5})
(which is valid for all non-zero replica field components
independently), the above instanton solution
$\f_{a}^{*}({\bf x})$, eq.(\ref{st3.6}), has additional 
degeneracy factor $2^{m}$.

The final step is the integration over fluctuations $\p_{a}({\bf x})$
at the background of the above instanton solution.  
Substituting $\f_{a}({\bf x}) = \f^{*}_{a}({\bf x}) + \p_{a}({\bf x})$,
and expanding the Hamiltonian up to the second order in 
$\p_{a}({\bf x})$, one has to perform the standard Gaussian
integration. These calculations, although 
slightly cumbersome, are quite straightforward
(for the details see Refs.\cite{grif-dos,grif-inst}). 
In the result for the Hessian factors one gets

\be
\lb{st3.15}
\(\det\hat T\)^{-1/2}_{n\to 0}  \simeq 
\exp\[  \fr{3m}{2(um-g)} g \psi_{o}^{2}\]
\ee
Comparing this with factor $\exp[-E(m)]$, where
$E(m)$ is the instanton energy, eq.(\ref{st3.12}),
we see, that under condition
\be
\lb{st3.16}
\t \gg \t_{g} = g^{2/(4-D)}
\ee
the contribution of fluctuations 
can be neglected. This is not surprising because eq.(\ref{st3.16})
is nothing else, but the Ginzburg-Landau criterion
which defines the temperature region away from $\T$,
where the critical fluctuations are irrelevant.
On the other hand, it has to be stressed that
in the close vicinity of $\T$ (at $\t \leq \t_{g}$), 
where the critical fluctuations
{\it are} relevant, the Gaussian 
approximation used for obtaining
the result, eq.(\ref{st3.15}), can not be valid anymore,
and to derive the corresponding fluctuations contribution
one would have to start some kind of RG procedure
which would properly take into account non-Gaussian
interactions. Thus, the above result for the 
fluctuations contribution, eq.(\ref{st3.15}), either
can be considered as the small correction (at 
$\t \gg \t_{g}$), or otherwise (if it is not small),
it is not valid (at $\t \leq \t_{g}$).

Thus, substituting the value of the instanton
energy, eq.(\ref{st3.12}), its size $R = \t^{-1/2}$
as well as its degeneracy factor $2^m$ into the series, 
eq.(\ref{st2.8}), we get

\be 
\lb{st3.17}
F_{G}  \; \simeq \;
- V \t^{D/2} \sum_{m=[g/u]+1}^{\infty} \fr{(-1)^{m-1}}{m} \; 2^{m}  \;
\exp\[-E_{0}(D) \fr{m}{um-g} \t^{(4-D)/2} \]
\ee
The exact summation of this series seems to be
rather tricky problem, but with the exponential accuracy
it can be estimated in a very simple way.
One can easily see that in the limit of weak disorder,
at $u \ll g$, the leading contribution in this summation
comes from the region $m \gg g/u \gg 1$ 
(where the exponential
factor in eq.(\ref{st3.17}) becomes $m$-independent)
and this contribution is
\be
\lb{st3.18}
 F_{G} \sim \exp\(-E_{0}(D) \fr{\t^{(4-D)/2}}{u} \)
\ee
We see that obtained off-perturbative (Griffith-like)
part of the free energy, as the function of the 
disorder parameter in the limit $u \to 0$, has the form 
of the essential singularity. 
Note again that this contribution
exists only in dimensions $D < 4$. As discussed above,
at $D \to 4$ (the upper critical dimensions),
the dimensionless instanton energy factor $E_{0}(D)$
approaches the finite universal limiting
value $E_{*} \simeq 26.3$.
In our world, in three dimensions, $E_{0}(D=3) \simeq 18.9$.

\section{Random field Ising model}

To study the off-perturbative effects in the 
random field Ising model we are going to use 
again the Ginzburg-Landau continuous representation:
\be
\lb{st4.1}
H = \I \Biggl[ \2 \(\n\f({\bf x})\)^{2}  
   + \2 \t \f^{2}({\bf x})  
   + \4 g \f^{4}({\bf x}) - h(x) \f({\bf x}) \Biggr] 
\ee  
Here the random function $h({\bf x})$ is described by the 
Gaussian distribution,
\be
\lb{st4.2}
P[h({\bf x})] = p_{0} 
\exp \Biggl( -\frac{1}{2h_{0}}\I h^{2}({\bf x}) \Biggr) \ ,
\ee
where $h_{0}$ is the small parameter which describes 
the effective strength of the random field, 
and $p_{0}$ is the normalization constant. 
Unlike the random temperature
model, considered in the previous section, here
we are going to consider 
the system in the low-temperature ferromagnetic phase
(supposing that the dimensionality $D$ is such that
this phase exists),
so that the reduced temperature 
parameter $\t$ will be taken to be negative,
$\t = -|\t|$. 
Again, we will place the system away from the critical
point, assuming that the absolute value $|\t|$
is not too small. As usual, to avoid the effects
of the critical fluctuations (in dimensions $D < 4$)
we impose the condition $|\t| \gg g^ {2/(4-D)}$.

Let us suppose that in the absence of the random fields
the ferromagnetic ground state of 
the system, eq.(\ref{st4.1}), is "up". This state 
(at the mean-field level) is characterized by the order parameter
\be
\lb{st4.3}
\f_{0} = +\sqrt{\fr{|\t|}{g}}
\ee
and the energy density 
\be
\lb{st4.4}
\e_{0} = - \fr{\t^{2}}{4 g}
\ee
In the usual perturbative approach 
the effects produced by the random field term of the 
Hamiltonian, eq.(\ref{st4.1}), together with the 
thermal fluctuations could be calculated 
in the systematic way in terms of the RG procedure 
(see e.g. \cite{RF-general} and references therein).
This approach is designed to take into account
all degrees of freedom at scales less that the correlation 
length, $\R(\t) \sim |\t|^{-1/2}$.

\begin{center}
\begin{figure}[h]
\includegraphics[scale=0.30]{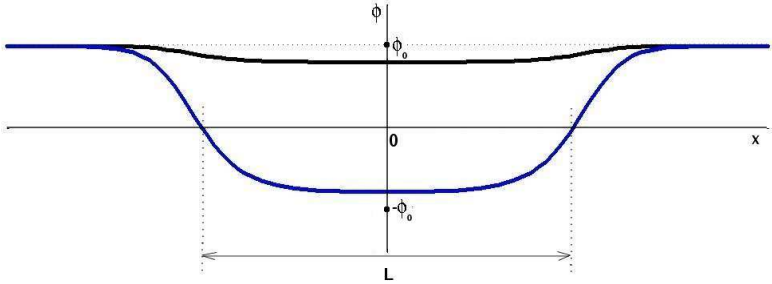}
\caption[]{Two alternative 
local minima field configurations
in a spatial "island" of the linear size $L$, where
the average value of the random field $h$ is negative
and its absolute value is not too small.}
  \label{fig5}
\end{figure}
\end{center}

On the other hand, at scales bigger that the correlation 
length we can observe completely different type of 
thermal excitations.
Let us consider a spatial island of the linear size $L$,
where the average value of the field $h$ is negative and its 
absolute value is not too small. Then, in addition to the 
state "up" (with slightly modified value of the order 
parameter), another local minimum with orientation
"down" can exist in this island (Fig.5). To be stable,
the gain in the energy due to the interaction with
the field, 
\be
\lb{st4.5}
E_{h} \; \sim \; - L^{D} |h| \f_{0}
\ee
should overrun the loss of energy due to the creation of the
domain wall,
\be
\lb{st4.6}
E_{d.w.} \; \sim \;  L^{D-1} \fr{\f^{2}_{0}}{\R}
\ee
Thus, such double-state situation in the considered island
is created provided
\be
\lb{st4.7}
|h| \; > \; \fr{\f_{0}}{L \R} \; \sim \; 
          \fr{|\t|}{L \sqrt{g}} \; \equiv \; h_{c}
\ee	  
According to eq.(\ref{st4.2}) the probability to find
an island of the size $L$ with the average value of the field
$h$ is
\be
\lb{st4.8}
P(L,h) \; \sim \; \exp\[-\fr{h^{2}}{2 h_{0}^{2}} \; L^{D} \]
\ee
Then the contribution to the free energy
of such rare "flipped" states can be estimated by their
probability:
\bq
\lb{st4.9}
F_{G} &\sim& \int_{\R}^{\infty} dL \int_{h_{c}}^{\infty} dh
         \exp\[-\fr{h^{2}}{2 h_{0}^{2}} \; L^{D} \] 
\nn
\nn	 
     &\sim& \int_{\R}^{\infty} dL
          \exp\[ -(const) \fr{\t^{2}}{h_{0}^{2} g} \; L^{D-2} \]
\nn
\nn	  
     &\sim& \exp\[-(const) \fr{\t^{\fr{6-D}{2}}}{h_{0}^{2} g} \]
\eq	 
Note that to obtain this result
in the above integration over $L$ the dimensionality of the 
system $D$ must be bigger than two (otherwise the integral
will become divergent). This is nothing else but 
slightly modified version of the good old Imri-Ma
arguments\cite{Imry-Ma} which tells that in dimension $D\leq 2$
flipping of magnetizations in big spatial islands
can become energetically favorable, which indicate 
the instability of the global ferromagnetic state.
Here we assume that the ferromagnetic state is stable,
and we see that rare off-perturbative flipping
excitations produce non-analytic contribution
to the free energy, which in the limit $h_{0} \to 0$
has the form of essential singularity.

\vspace{5mm}

Now we are going to re-derive the above prediction,
eq.(\ref{st4.9}), in terms of much more rigorous
systematic procedure described in section II. 
Coming back to the original Hamiltonian, eq.(\ref{st4.1}),
after the Gaussian averaging of the replicated partition
function over the random function $h({\bf x})$,
one obtains the replica Hamiltonian
\be
\lb{st4.10}
H_{n}\[{\boldsymbol\f}\] \; = \;
 \I \Biggl[ \2 \sum_{a=1}^{n}\(\n\f_{a}\)^{2} 
      - \2 |\t| \sum_{a=1}^{n} \f_{a}^{2} 
      + \4 g \sum_{a=1}^{n} \f_{a}^{4} 
      - \2 h_{0}^{2} \sum_{a,b=1}^{n}  \f_{a} \f_{b}
   \Biggr]
\ee
The saddle-point configurations of the fields
$\f_{a}(x)$ are defined by the equations
\be
\lb{st4.11}
-\D\f_{a}({\bf x}) - |\t| \f_{a}({\bf x}) + g \f_{a}^{3}({\bf x}) 
-h_{0}^{2} \(\sum_{b=1}^{n} \f_{b}({\bf x})\) = 0
\ee  
Below we are going to demonstrate that besides the 
obvious (replica symmetric) ferromagnetic solution 
$\f_{a}({\bf x}) = \sqrt{|\t|/g}$ these equations
have non-trivial localized in space
instanton-like solutions with the RSB two-groups
structure:

\be
\lb{st4.12}
\f_{a}^{*}({\bf x}) = \left\{ \begin{array}{ll}
\sqrt{\fr{|\t|}{g}} \; \psi_{1}({\bf x}\sqrt{|\t|}) &\mbox{for $a=1,...,m$}
		 \\
		 \\
\sqrt{\fr{|\t|}{g}} \; \psi_{0}({\bf x}\sqrt{|\t|}) &\mbox{for $a=m+1,...,n$}
                            \end{array}
                            \right.
\ee
Substituting these rescaled fields
 into the saddle-point eqs(\ref{st4.11})
and into the Hamiltonian, eq.(\ref{st4.10}), we find that
(in the limit $n\to 0$) the instanton configuration 
$\{\psi_{1}({\bf z}),\psi_{0}({\bf z})\}$ (where ${\bf z}={\bf x}\sqrt{|\t|}$)
is defined by the two equations
\bq
\lb{st4.13}
-\D\psi_{1} - \psi_{1} + \psi_{1}^{3} - \lm(m) \(\psi_{1}-\psi_{0}\) &=& 0
\nn
\nn
-\D\psi_{0} - \psi_{0} + \psi_{0}^{3} - \lm(m) \(\psi_{1}-\psi_{0}\) &=& 0
\eq  
and its energy is
\be
\lb{st4.14}
E(m) = m \fr{|\t|^{2-D/2}}{g}
\int d^{D} z \Biggl[ \2 \[(\n\psi_{1})^{2}-(\n\psi_{0})^{2}\] 
	        -\2 \[\psi_{1}^{2} -\psi_{0}^{2}\]
	        +\4 \[\psi_{1}^{4} - \psi_{0}^{4}\]
		-\2 \lm(m) \[\psi_{1} -\psi_{0}\]^{2}
		\Biggr] 
\ee
where
\be
\lb{st4.15}
\lm(m) \; = \; \fr{h_{0}^{2} m}{|\t|}
\ee
We are looking for the localized in space
(spherically symmetric) solutions of the eqs.(\ref{st4.13}),
such that the two functions $\psi_{1}(r)$ and $\psi_{0}(r)$
(where $r = |{\bf z}|$)
are different from each other in a finite region of space, 
and at large distances they both sufficiently quickly 
approach the same value $\psi=1$,
so that the integral in eq.(\ref{st4.14}) will be converging.
Simple analysis of the structure of the 
"potential energy" 
\be
\lb{st4.16}
U(\psi_{1},\psi_{0}) \; = \; 
	        -\2 \[\psi_{1}^{2} -\psi_{0}^{2}\]
	        +\4 \[\psi_{1}^{4} - \psi_{0}^{4}\]
		-\2 \lm \[\psi_{1} -\psi_{0}\]^{2}
\ee
shows that until the parameter $\lm$ is small
(so that the last coupling term in the above expression 
is just a small correction), the potential 
$U(\psi_{1},\psi_{0})$
has 9 saddle-points (in the vicinity of the points
$(0;0), (0;\pm 1), (\pm 1; 0), (\pm 1, \pm 1)$
and $(\pm 1;\mp 1)$). In this situation
the two fields $\psi_{1}$ and $\psi_{0}$ are 
effectively independent, and no instanton-like
solutions described above can exist.
When increasing the parameter $\lm$,
starting from 
\be
\lb{st4.17}
\lm > \lm_{c} \simeq 0.23
\ee 
only 5 saddle points of the potential $U(\psi_{1},\psi_{0})$ 
remains in the plain $(\psi_{1};\psi_{0})$. They have
coordinates: $(0;0), (\pm 1;\pm 1)$ and 
$(\pm \psi_{1}^{*}(\lm);\pm \psi_{0}^{*}(\lm))$, where
$0 < \psi_{1,0}^{*} < 1$ (in particular, 
$\psi_{1}^{*}(\lm_{c}) \simeq 0.17$
and $\psi_{0}^{*}(\lm_{c}) \simeq 0.90$). It is crucial that
at the points $(\pm \psi_{1}^{*};\pm \psi_{0}^{*})$
the potential $U(\psi_{1},\psi_{0})$ has the 
{\it maxima}. It is due to the existence of these maxima 
that at $\lm > \lm_{c}$ the instanton solutions become
possible. 

Let us consider the limit  $\lm(m) \gg 1$, or
\be
\lb{st4.18}
m \; \gg \; m_{c} \; = \; \[\lm_{c} \fr{|\t|}{h_{0}^{2}}\] + 1
\ee
In this limit, 
according to eqs.(\ref{st4.13}), 
the two fields $\psi_{1}$ and $\psi_{0}$ must be
close to each other. Redefining,
\bq
\lb{st4.19}
\psi_{1}(r) \; &=& \; \psi(r) + \fr{1}{\lm} \chi(r)
\nn
\nn
\psi_{0}(r) \; &=& \; \psi(r) - \fr{1}{\lm} \chi(r)
\eq 
in the leading order in $\lm^{-1}$ instead of
eqs.(\ref{st4.13}) we get much more simple equations:
\bq
\lb{st4.20}
-\D\psi - \psi + \psi^{3} - 2 \chi &=& 0
\nn
\nn
-\D\chi  + (3 \psi^{2} - 1) \chi &=& 0
\eq  
which {\it contain no parameters}. For the energy of
the configurations described  by the two fields
$\psi(r)$ and $\chi(r)$ instead of eq.(\ref{st4.14})
(again, in the leading order in $\lm^{-1}$) 
we find the value, which does not depend on
the summation parameter $m$,
\be
\lb{st4.21}
E \; = \; \fr{|\t|^{\fr{6-D}{2}}}{h_{0}^{2} g} \; E_{0}(D)
\ee
where 
\be
\lb{st4.22}
E_{0}(D) \; = \; 
\int d^{D} z \[ (\n\psi)(\n\chi) 
             + (\psi^{3} - \psi)\chi - \chi^{2} \]
\ee
is the universal quantity which depends only on the 
dimensionality of the system.

It turns out that in dimensions $D < 3$,
the system of eqs.(\ref{st4.20}), 
indeed has smooth instanton-like
spherically symmetric solution which
has finite and positive energy $E_{0}(D)$. 
Within the limited spatial region 
$r \leq r_{c} \sim 1$,
the values of the fields
$\psi(r)$ and $\chi(r)$  
are finite and of the
order of their values at the origin, 	  
$\psi_{0} \sim 1$ and $\chi_{0} \sim 1$. 
On the other hand,
at $r \gg r_{c}$ the function 
$\psi(r)$ exponentially
quickly approaches 1, while the function
$\chi(r)$ exponentially tents to zero.
The illustration of the instanton solution
in the dimension $D = 2.9$ is given in Fig.6, where 
$\psi_{0} \simeq -0.818$, $\chi_{0} \simeq -0.284$

\begin{center}
\begin{figure}[h]
\includegraphics[scale=0.40]{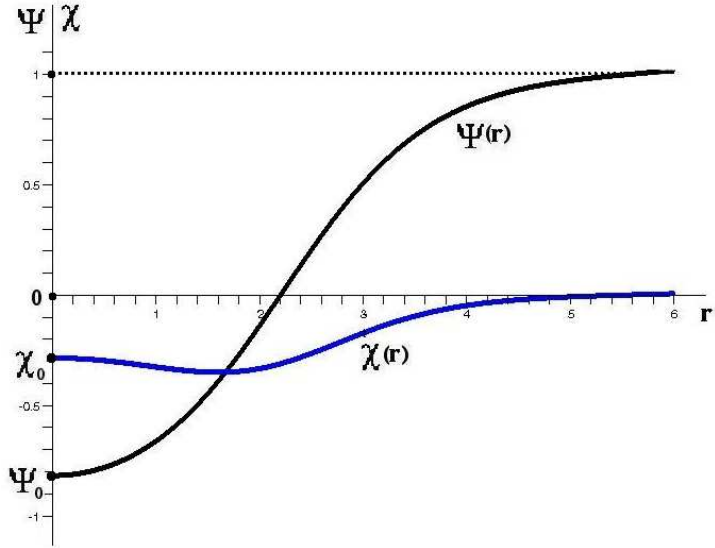}
\caption[]{The instanton solution $\psi(r)$, $\chi(r)$ 
of eqs.(\ref{st4.20})
in dimensions $D=2.9$}
  \label{fig6}
\end{figure}
\end{center}

Fig.7 demonstrates the corresponding "trajectories" of the
instanton solutions in the plane $(\psi,\chi)$ at various 
values of the dimension.
As the dimension $D$ approaches the
value $D_{c}=3$ from below, the starting values
$\psi_{0} \to -1$ and $\chi_{0} \to 0$. 
Above three dimension the instanton solution
disappears. Thus we have to conclude that $D=3$ is the upper 
critical dimension for the 
considered Griffiths phenomena in the RFIM, in agreement with the 
earlier suggestion \cite{RF-grif-dos1,RF-grif-dos2} 
as well as with the recent studies
of similar instanton-like configurations in the presence of
external magnetic field \cite{mueller-silva}
\begin{center}
\begin{figure}[h]
\includegraphics[scale=0.40]{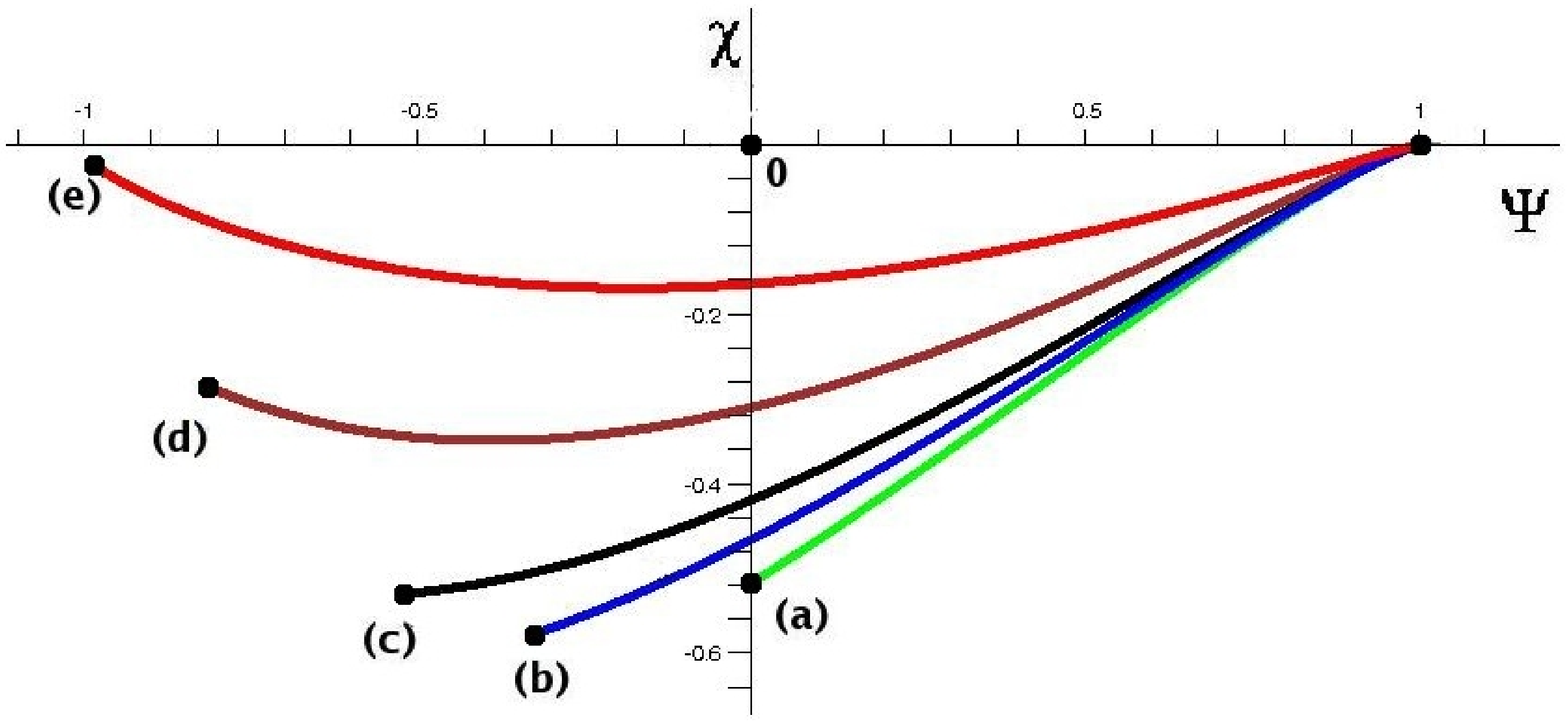}
\caption[]{The instanton solution "trajectories"
in the plane ($\psi, \chi$) in various dimensions:\\
(a) $D=2$,    $\; \psi_{0}\simeq  0.001, \;\chi_{0}\simeq -0.5173$;
(b) $D=2.5$,  $\; \psi_{0}\simeq -0.325, \;\chi_{0}\simeq -0.5776$;
(c) $D=2.7$,  $\; \psi_{0}\simeq -0.520, \;\chi_{0}\simeq -0.5300$;
(d) $D=2.9$,  $\; \psi_{0}\simeq -0.818, \;\chi_{0}\simeq -0.2844$;
(e) $D=2.92$, $\; \psi_{0}\simeq -0.988, \;\chi_{0}\simeq -0.0212$.}
  \label{fig7}
\end{figure}
\end{center}

Let us come back to the general expression for the
off-perturbative part of the free energy,
the series, eq.(\ref{st2.8}), 
where the summation over $m$ starts now from
$m = m_{c}$, eq.(\ref{st4.18}). 
Noting that the
instanton energy $E(m)$ is the {\it decreasing} 
function of $m$, we can conclude that with the exponential
accuracy this {\it converging} series can be estimated
by its asymptotic part at $m \gg m_{c}$.
Thus, substituting here the value of the instanton energy, 
eq.(\ref{st4.21}), (and neglecting the critical
fluctuations), with the exponential accuracy
we get the result
\be
\lb{st4.24}
F_{G} \; \sim \; 
\exp\[- E_{0}(D) \fr{|\t|^{\fr{6-D}{2}}}{h_{0}^{2} g} \]
\ee
which perfectly agrees with the "hand-waving"
estimate, eq.(\ref{st4.9}).
This non-analytic in $h_{0}$ contribution, 
which has the form of essential singularity
at $h_{0} \to 0$, is valid only in dimensions
$D < 3$, and at temperatures not too close to the 
critical point, at $|\t| \gg g^{2/(4-D)}$. The dimension
$D=3$ is marginal for this kind of phenomena. 
Therefore the investigation of the Griffith-like 
contributions in the three-dimensional RFIM requires 
much more discreet analysis.

\section{Discussion}

In this paper the systematic method for computing 
off-perturbative thermodynamic contributions 
in disordered systems has been proposed.
It has been tested on two the most popular
classical statistical systems containing
quenched disorder: the random temperature and 
the random field Ising models. In both cases
the off-perturbative contributions as the 
functions of the parameters which describe
the strength of the disorder have the form 
of the essential singularities, eqs.(\ref{st3.18}), 
(\ref{st4.24}). 

Of course, thinking about possible
experimental or numerical tests,
the validity
of the obtained results is rather limited.
On one hand, since the consideration has been 
done in terms of the continuous
Ginsburg-Landau Hamiltonian, one
has to place the system sufficiently close to the
phase transition point, so that the correlation 
length would be large compared to the 
lattice spacing. On the other hand,
since the present study completely neglects
the critical fluctuations, the system
has to be sufficiently far from the 
critical point. Formally, in terms of the
Ginsburg-Landau Hamiltonian, these two requirements
can be easily satisfied: it is sufficient
to demand that (1) the coupling parameter $g$
is small, and (2) the reduced temperature
parameter $\t$ is bounded by the condition
$g^{2/(4-D)} \ll |\t| \ll 1$ (where $D$ is the 
system dimensionality, $D < 4$). 
Besides, the strength of the disorder must be
small: $u \ll g$ in the random temperature model,
and $h_{0} \ll \sqrt{|\t|}$ in the random field one.

On the other hand, keeping in mind more general 
perspectives, the proposed approach, 
in my view, may open the way to 
study the nature of the phase transitions
in the considered systems. 
It is generally believed that it is the 
off-perturbative states which makes the study
of the phase transition in the random field
Ising model so difficult\cite{RF-general}. 
It is remarkable, that according to the present
study, quite similar off-perturbative 
contributions are also present in the
random temperature Ising systems
where, at least at the qualitative level,
the nature of the phase transition  
was traditionally believed  to be well
understood (see e.g. \cite{RG-disorder,dos-book}).
This supports recent suspects\cite{critical-rsb} 
that the off-perturbative
effects could be quite relevant for the critical properties
of the disordered ferromagnetic systems.

Another interesting observation is that 
in terms of the considered off-perturbative
contributions, the situation, when approaching $\T$ 
from above and from below, looks totally
asymmetric both in the random field and in the random
temperature systems. The considered effects
are present in the ferromagnetic phase, at $T < \T$,
of the random field model, while they are
{\it absent} in its paramagnetic phase at $T > \T$. 
On the other hand, the  situation
in the random temperature model is similar,
although "reversed": the off-perturbative contributions
are present in the paramagnetic phase, at $T > \T$,
while they are absent in the ferromagnetic phase,
at $T < \T$. May be it is this asymmetry,
which makes the nature of the  phase transitions
in these systems to be so non-trivial?

One more qualitative observation is that 
according to the present study, the off-perturbative
contributions are {\it absent} in the
random temperature model at dimensions $D > 4$,
and in the random field model at dimensions $D > 3$.
As for the random temperature
systems, this is not surprising:
all the previous studies were definite that 
at $D > 4$ the disorder is irrelevant for the 
phase transition. What does this mean
for the random field model is much less clear,
because here it is well established that
its upper critical dimensionality
is equal to 6 (the dimensionality above which
the critical behavior is described by the Gaussian
theory, and the presence of the random
fields is irrelevant for the phase transition).
Well, of course, the absence of the off-perturbative
contributions in dimensions $3 < D \leq 6$
does not mean, that the random
fields are irrelevant. Probably it indicates that
here all the random field effects can be taken
into account in the framework of 
the perturbative RG procedures.

To answer all the above questions
(as well as many other important questions
which were not formulated here),
the only thing which remains to be done
is to find the way to overcome the 
Ginzburg-Landau limitation
$|\t| \gg g^{2/(4-D)}$, and to take the limit $\t \to 0$.
To do that one has just to formulate 
a theory which would properly take into account 
the critical fluctuations on top of the
instanton-like configurations described
in this paper.

\vspace{15mm}

{\bf Acknowlegments}

\vspace{5mm}

The author is grateful to Markus Mueller and Alessandro Silva
for their quite helpful comments and advises concerning the 
numerical solutions for the instanton configurations 
in the random field Ising model.


\begin{thebibliography}{99}



\bibitem{harris} A.B.Harris, 
               Phys.Rev. {\bf B12}, 203 (1975)
	       
\bibitem{RG-disorder}  
         A.B.Harris and T.C.Lubensky, 
	      Phys.Rev.Lett., {\bf 33}, 1540 (1974);
          D.E.Khmelnitskii, 
	      ZhETF (Soviet Phys. JETP) {\bf 68}, 1960 (1975);
          G.Grinstein and A.Luther, 
	      Phys.Rev. {\bf B 13}, 1329 (1976);
	  Vik.S.Dotsenko and Vl.S.Dotsenko,
	      Adv.Phys. {\bf 32}, 129 (1983)
	      

\bibitem{grif} R.Griffiths, 
               Phys.Rev.Lett. {\bf 23}, 17 (1969)
	       

\bibitem{grif-stat}
         M.Wortis, 
               Phys.Rev. {\bf B10}, 4665 (1974);
         A.B.Harris, 
               Phys.Rev. {\bf B12}, 203 (1975);
         Y.Imry,  
               Phys.Rev. {\bf B15}, 4448 (1977);
         A.J.Bray and D.Huifang, 
               Phys.Rev. {\bf B40}, 6980 (1989);
         J.J.Ruiz-Lorenzo, 
              J.Phys.A: Math.Gen., {\bf 30}, 485 (1997)


\bibitem{zeros-cardy} 
         J.L.Cardy and A.J.McKane, 
              Nucl.Phys. B {\bf 257} [FS14] 383 (1985)

\bibitem{grif-dos} 
         Vik.S.Dotsenko, 
	       Preprint cond-mat/0505233.

\bibitem{grif-dynam} 
         M.Randeria, J.P.Sethna and R.Palmer,
	      Phys.Rev.Lett. {\bf 54}, 1321 (1985);         
	 A.T.Ogielski, 
	       Phys.Rev. {\bf B32}, 7384 (1985)
	 A.J.Bray, 
              ibid. {\bf 59}, 586 (1987);
	 D.Dhar, M.Randeria and J.P.Sethna,
	      Europhys.Lett. {\bf 5}, 485 (1988)
	 A.J.Bray and G.J.Rodgers,
	      Phys.Rev. {\bf B38}, 9252 (1988)
	 S.Calborne and A.J.Bray,
	      J.Phys. {\bf A22}, 2505 (1989)  
         S.Jaine,
	      Physica {\bf A218}, 279 (1995)
         F.Cesi, C.Maes and F.Martinelli,
	       Comm.Math.Phys. {\bf 188}, 135 (1997);
	      

\bibitem{grif-quant}
         A.J.Millis, D.K.Morr, and J.Schmalian,	      
               Phys.Rev. {\bf B 66}, 174433 (2002);
	 V.Dobrosavljevic and E.Miranda,
	       Phys.Rev.Lett. {\bf 94}, 187203 (2005) 

\bibitem{grif-parisi} 
         G.Parisi,
	       in {\it Recent Advances in the Field Theory
	       and Statistical Mechanics}, Les Houches 1982
	       (Amsterdam: North-Holland)


\bibitem{RF-grif-dos1}
         G.Parisi and Vik.S.Dotsenko
	       J.Phys. {\bf A25}, 3143 (1992)


\bibitem{RF-grif-dos2}	
         Vik.S.Dotsenko
	       J.Phys. {\bf A27}, 3397 (1994)  
	       
\bibitem{vrsb}
         Vik.S.Dotsenko and M.Mezard, 
	       J.Phys. {\bf A30}, 3363 (1997)	            

\bibitem{dos-book}   
           Vik.S. Dotsenko, 
	      {\it Introduction to the Replica Theory
              of Disordered Statistical Systems}
              (Cambridge University Press, 2001).

\bibitem{grif-inst} 
           Vik.S.Dotsenko, 
              J.Phys. {\bf A32}, 2949 (1999)

	       
	     	      
\bibitem{MPV-book}
          M.M\'{e}zard, G.Parisi, and M.A.Virasoro,
              {\it Spin glass theory and beyond} 
	      (World Scientific, Singapore, 1987);

\bibitem{critical-rsb}
  Vik.S.Dotsenko, B.Harris, D.Sherrington and R.Stinchcombe,
              J.Phys. {\bf A28}, 3093 (1995);
  Vik.S.Dotsenko, Vl.S.Dotsenko, M.Picco and P.Pujol,
              Europhys.Lett. {\bf 32}(5), 425 (1995);	      
  G.Tarjus and Vik.S.Dotsenko,
	      J.Phys. {\bf A35}, 1627 (2002)

	      
\bibitem{inst} 
          J.Zinn-Justin, 
	      {\it Quantum Field Theory 
	      and Critical Phenomena},
              Clarendon Press (Oxford 1989, 
	      third ed. 1996)
	      
  
\bibitem{RF-general} 
     {\it Spin glasses and random fields},
     edited by A.P.Young
     (World Scientific, Singapore 1998)

\bibitem{Imry-Ma}
        Y.Imry and S.-K.Ma,
	   Phys.Rev.Lett. {\bf 35}, 1399 (1975)	      

\bibitem{mueller-silva}  M.Mueller and A.Silva,
        "Instanton analysis of hysteresis in the 
	3D Random Field Ising Model", Preprint cond-mat/0505048	         


\end{thebibliography}
\end{document}